\documentstyle[epsfig,rotating]{article}
\def\Journal#1#2#3#4{{#1} {\bf #2}, #3 (#4)}

\def\NIMA{{\em Nucl. Instrum. Methods} A}

\def\NPBP{{\em Nucl. Phys.} B (Proc. Suppl.)}
\def\PLB{{\em Phys. Lett.}  B}
\def\PRL{\em Phys. Rev. Lett.}
\def\PRD{{\em Phys. Rev.} D}

\def\RPP{\em Rep. Prog. Phys.}
\def\PPNP{\em Prog. Part. Nucl. Phys.}
\def\RMP{\em Rev. Mod. Phys.}
\def\APJ{\em ApJ }

\def\ra{\rightarrow}
\def\be{\begin{equation}}
\def\ee{\end{equation}}
\def\bea{\begin{eqnarray}}
\def\eea{\end{eqnarray}}
\newcommand{\nue}{neutrino }
\newcommand{\nues}{neutrinos }
\newcommand{\hdm}{hot dark matter }
\newcommand{\hdmp}{hot dark matter. }

\newcommand{\cdmp}{cold dark matter. }
\newcommand{\dm}{dark matter }
\newcommand{\nosz}{neutrino oscillations }
\newcommand{\noszp}{neutrino oscillations. }
\newcommand{\osz}{oscillation }
\newcommand{\oszsp}{oscillations. }
\newcommand{\oszs}{oscillations }

\newcommand{\atm}{atmospheric neutrinos }
\newcommand{\ssm}{see-saw-mechanism }
\newcommand{\adn}{almost degenerated neutrinos }
\newcommand{\delm}{\mbox{$\Delta m^2$} }
\newcommand{\me}{\mbox{$m_{\nu_e}$} }
\newcommand{\mmu}{\mbox{$m_{\nu_\mu}$} }
\newcommand{\mtau}{\mbox{$m_{\nu_\tau}$} }
\newcommand{\nel}{\mbox{$\nu_e$} }
\newcommand{\nmu}{\mbox{$\nu_\mu$} }
\newcommand{\ntau}{\mbox{$\nu_\tau$} }
\newcommand{\sint}{\mbox{$sin^2 2\theta$} }
\newcommand{\lbls}{long baseline experiments }
\newcommand{\lbl}{long baseline experiment }
\begin{document}
\title{NEUTRINO OSCILLATIONS AND DARK MATTER
\footnote{to appear in Proc. Workshop {\it Aspects of Dark Matter in
Astro- and Particle Physics (Dark 96)}, Heidelberg Sep. 1996, Germany}}
\author{ K. ZUBER 
\smallskip\\
Lehrstuhl f\"ur Exp. Physik IV, Universit\"at Dortmund,\\ 
Otto Hahn Str. 4, 44221 Dortmund,
Germany}
\date{}
\maketitle
\begin{abstract}
The significance of light massive \nues as hot dark matter is outlined.
The power of \nue \osz experiments with respect to detect such \nues in the
eV-region
is
discussed. Present hints for \nosz in solar, atmospheric and LSND data are
reviewed as well as future experiments and their potential.
\end{abstract}
\section{Introduction}
Most astrophysical models which describe large-scale structure and the cosmic
microwave background consistently end up with a mixture of \hdm and \cdmp The
most obvious
candidates for \hdm are stable neutrinos with masses in the eV-region. Their
number density is given by
\be 
n_{\nu_i} = \frac{3}{11} \cdot (\frac{g_{\nu_i}}{2}) \cdot
n_{\gamma_0}
\ee  
resulting in a contribution to the mass density of (assuming $g_{\nu_i} = 2$) 
\be
\rho_{\nu_i} = \sum m_{\nu_i} n_{\nu_i} \Rightarrow \sum m_{\nu_i} = 94 \Omega_{\nu}
h^2 eV
\ee
If as suggested $\Omega_{\nu} h^2$ is about 0.2 - 0.3 any \nue in the eV region
contributes significantly.\\
Two groups of models for \nue masses emerged over the past years to explain the
present status of \nue observations.
The first one is the ''classical'' quadratic \ssm resulting in a strong scaling
behaviour of the
neutrino masses like
\be
\me : \mmu : \mtau \propto m_u^2 : m_c^2 : m_t^2
\ee
where $m_u,m_c$ and $m_t$ are the corresponding quark masses. Because of the large
mass difference between the involved quarks only one \nue can act as a \dm
candidate, for example $\ntau$. Recently another type of \ssm was coming up
resulting in more or less \adn \cite{rabi}. This of course offers the chance that
two or
more \nues can act as \hdmp Indeed some astrophysical models favour scenarios with
2 \nues as
\hdm \cite{joel}. 
The present direct limits for \nue masses are \cite{nmass}:
\begin{center}
\begin{tabular}{rl}
\me $<$ &15 eV \\
\mmu $<$ & 170 keV \\
\mtau $<$ & 18.2 MeV
\end{tabular}
\end{center}
\medskip
The actual results for $\me$ taken from tritium beta decay experiments are 
better and
give a limit of 3.5 eV, but face the problem of negative $m^2$-values.
Another bound valid only for
Majorana neutrinos results from double beta decay and is given by \cite{heimo}
\be
\langle \me \rangle < 0.5 eV
\ee
What can be seen from the above is, that a direct kinematic test of \mmu and
\mtau or even \me in the eV or even sub-eV-region is impossible, the only
possibility to explore this region is \noszp
\section{Neutrino oscillations}
As in the quark sector also for neutrinos the mass eigenstates need
not to be the
same as the flavour eigenstates offering the possibility of oscillations.
In the case of two flavours the mixing can be described by
\be
{\nu_e \choose \nu_{\mu}} = \left( \begin{array}{cc}
 cos \theta & sin \theta \\
- sin \theta & cos \theta
\end{array} \right)
{\nu_1 \choose \nu_2}
\ee
While $\sint$ describes the amplitude of the
oscillation, $\delm = m_2^2 -m_1^2$ determines the oscillation length,
characterising a full
cycle
of \osz between two flavours. 
In practical units the oscillation length is given by
\be
L_V = \frac{4\pi E \hbar}{\Delta m^2 c^3} =
2.48 (\frac{E}{MeV})(\frac{eV^2}{\Delta m^2}) \quad m
\ee
As can be seen \oszs do not allow an absolute mass measurement. Furthermore to
allow oscillations, the \nues are not allowed to be exactly degenerated.
Assuming the classical \ssm and $\ntau$ as \hdm candidate the
typical region to test is $\delm$ between 1-$10^3 eV^2$. In contrast to this
situation 
the almost degenerated scenario makes an investigation of the whole parameter
space for possible
\osz signals necessary.
\section{Present hints for \nosz}
\begin{figure}[hhh]
\begin{center}
\begin{tabular}{cc}
\epsfig{file=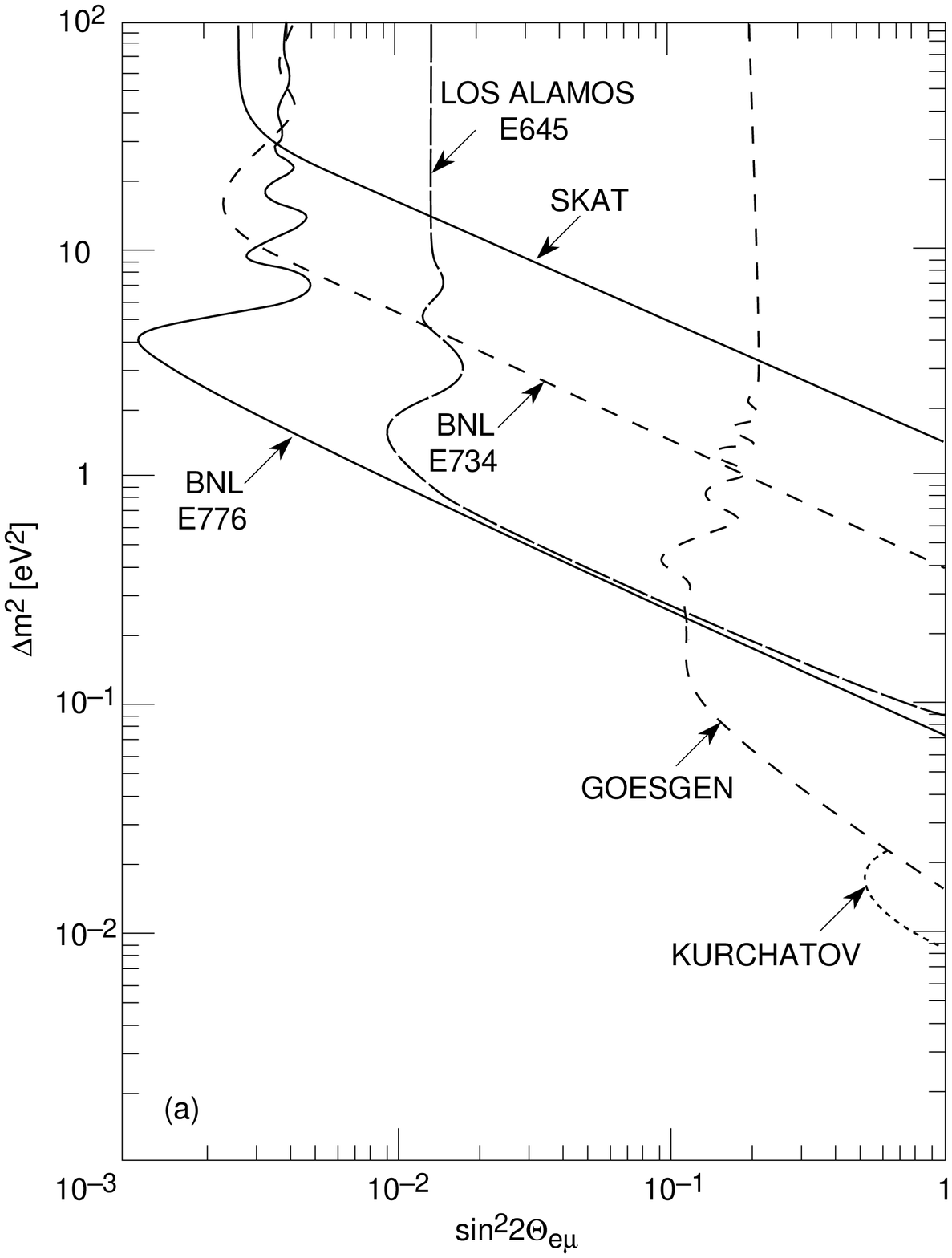,width=5.5cm,height=8cm} &
\epsfig{file=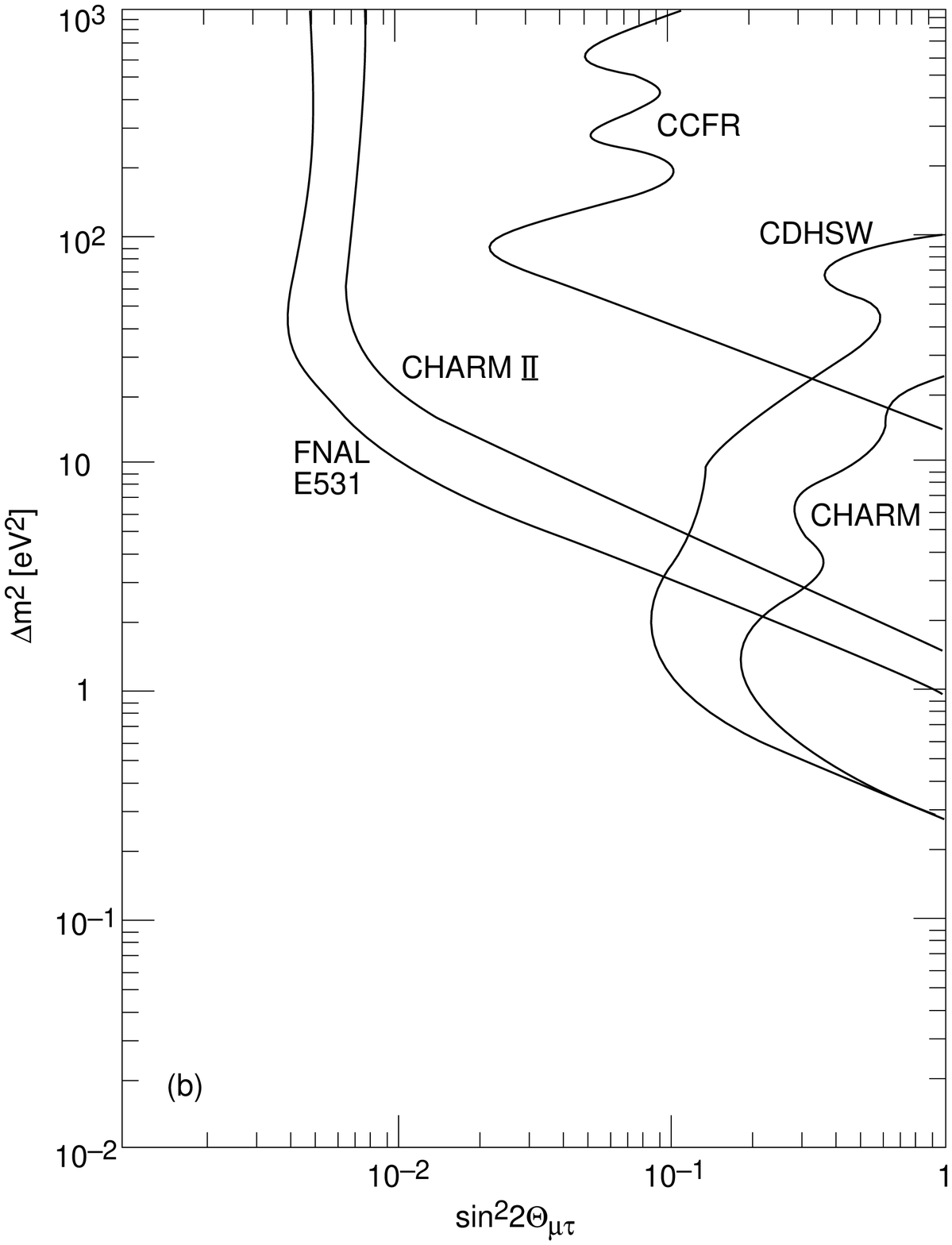,width=5.5cm,height=8cm}
\end{tabular}
\end{center}
\caption{\label{lego} \it Exclusion plots  on \nosz parameters from different
reactor and
accelerator experiments. Shown are \nel - \nmu (left) and \nmu - \ntau
(right) \oszsp The regions on the right side are excluded (from [5]).}
\end{figure}
\begin{figure}[hhh]
\begin{center}
\epsfig{file=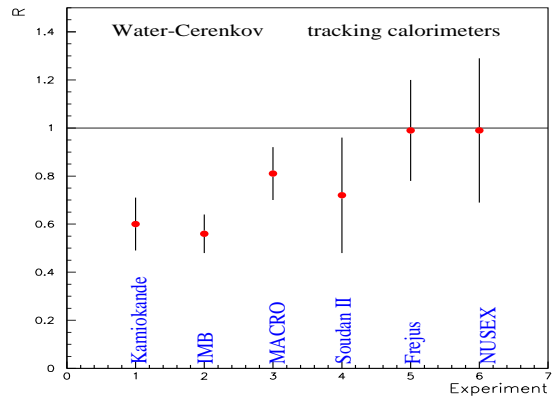,width=8cm,height=6cm}
\end{center}
\caption{\label{atres} \it Comparison of R as defined in eq.7 from different measrurements on
atmospheric
neutrinos. While the water Cerenkov detectors see a deficit from the expected
value of R = 1, most of the iron tracking calorimeters are more or less in
agreement with the expectation.} 
\end{figure}

Besides many experiments which did not find any evidence for \nosz (see fig.
1) there
are at present three fields of \nue physics which at least show hints for 
\noszp The first one is the long outstanding solar \nue problem, now showing up
in four different experiments, which have partly different energy thresholds. The
present status is shown in Tab.1.\\
Assigning the observed deficit to \nosz two solutions remain. First off all there exists the
possibility of vacuum \oszs with $\delm \approx 10^{-10} eV^2$ and large mixing
angles ($\sint \approx 1$) or, by considering \nosz in matter via the MSW-effect,
the results are two regions around $\delm \approx 10^{-5} eV^2$ with a
large-angle solution ($\sint \approx 1$) and a small angle solution ($\sint
\approx 10^{-2}$).\\
Another field where hints are seen is the one of atmospheric neutrinos. 
The production of pions and kaons in the atmosphere by cosmic ray primaries as
well as their decays lead
to predictions of observable fluxes of \nmu and $\nu_e$.
Because of absolute flux uncertainties people rely on the ratio R defined as
\be
R = \frac{(\mu /e)_{Data}}{(\mu /e)_{MC}}
\ee
\begin{center}
\begin{tabular}{|c|c|c|c|}
\hline
Experiment & Result & Prediction BP & Prediction TC \\
\hline
GALLEX [SNU]& $69.7 \pm 6.7 ^{+3.9}_{-4.5}$ & $131.5^{+6}_{-7}$ & $122.5 \pm
7$ \\
SAGE [SNU]& $69 \pm 10 ^{+5}_{-7}$ & $131.5^{+6}_{-7}$ & $122.5 \pm 7$
\\
\hline
Cl [SNU]& $2.56 \pm 0.16 \pm 0.14$ & $9.3^{+1.2}_{-1.4}$ & $6.4 \pm 1.4$  \\
\hline
Kamiokande & $2.80 \pm 0.19 \pm 0.33 $ & $42.4%
\pm 5.8 \% \cdot \Phi_t$ & $63.6 \% \pm 8.9 \% \cdot \Phi_t$ \\
\hline
\end{tabular}
\newline
\end{center}
\medskip
{\it Tab. 1: Comparison of experimental solar \nues results with predictions of
two solar model
calculations (BP \cite{bah}, TC \cite{tc}). The results for the gallium and
chlorine-experiments are given in SNU
(1 SNU = 10$^{-36}$ captures per target atom per second), while the Kamiokande result 
is given as flux in units of  $10^6 cm^{-2}s^{-1}$. The mentioned theoretical
values
correspond to the fraction of expected events. It should be mentioned that there
are
solar models which predict the right value for Kamiokande.}
\medskip\\
where $\mu$ expresses muon-like events and {\it e} electron-like events.
The present experimental values are (see also fig. \ref{atres}):
\begin{center}
\begin{tabular}{lrl}
Kamiokande:& R & = 0.60 $\pm$ 0.06 $\pm$ 0.05 \\
IMB:& R &= 0.56 $\pm$ 0.04 $\pm$ 0.04 \\
Soudan II:& R &= 0.72 $\pm 0.19 ^{+0.05}_{-0.07}$ \\
MACRO:& R &= 0.87 $\pm$ 0.05 $\pm$ 0.06 \\
Frejus:& R &= 0.99 $\pm$ 0.13 $\pm$ 0.08 \\
NUSEX:& R &= 0.99$^{+0.35}_{-0.25}$ 
\end{tabular}
\end{center}
A detailed analysis shows that it is a deficit of
muon \nues which is responsible for the lower R value in the water Cerenkov detectors. The
typical region to
describe this deficit via \nosz is at $\delm \approx 10^{-2} eV^2$ and $\sint
\approx 1$. 
A third hint comes from the LSND-experiment
having excess events
which can be explained by $\bar {\nel} \ra \bar {\nmu}$ - oscillations
\cite{lsnd}.
\section{Future checks on solar \nues }
To investigate the \osz hypothesis for solar \nues several experiments are
coming up. The first one is the Superkamiokande experiment running since 1. April
1996. Because of the 10-fold increase in fiducial volume, this detector will allow
a high statistics measurement of the $^8$B-flux. Another experiment which
will start data taking soon is a new radiochemical detector using 
in a first phase 100t $^{127}I$ with a threshold energy of 
$E_{Thr}$ = 789 keV. This experiment is installed in the Homestake-mine, close to
the Cl-experiment. A big step forward in testing the \osz hypothesis is the 
Sudbury Neutrino Observatory (SNO) which is by using D$_2$O able to measure
flavour-blind (neutral current) and flavour-sensitiv (charge current) via the
reactions
\be
\nu_e + d \rightarrow e^- + p + p \quad \mbox{and} \quad \nu + d \rightarrow \nu
+ p + n
\ee
It will be in operation in 1997. The present experimental situation of
solar \nues results requires more or less the absence of $^7$Be neutrinos. The
BOREXINO-experiment at Gran Sasso is especially designed to measure these
neutrinos \cite{borex}. A proposal for a real time pp-neutrino measurement is the
HELLAZ
experiment which could be online early next century.\\
A completly independent way to attack the solar neutrino problem is further
investigation of the
internal structure of the sun, especially with the help of helioseismology. Two new
projects, the
SOHO-satellite and the Global Oscillation Network Group (GONG) will give new
insights to this and will help to define the temperature and density profile of
the sun more accurate.
\section{Future checks on \atm }
The main impact in the near future on \atm will be Superkamiokande. It will allow
a
high statistics 
measurement of atmospheric neutrinos. The allowed parameter region can also be
checked by upcoming 
reactor experiments and \lbls with accelerators. Concerning reactor experiments
these
are the
CHOOZ experiment in France, which started data taking recently and the Palo Verde
experiment in the US, having a first module running by end of the year. Long
baseline experiments at accelerators will be discussed later.
\section{Future checks on LSND }
The allowed parameter range of LSND is already in some conflict with past or
ongoing experiments. A large part of their allowed range seems to be excluded
by the BNL E776, CCFR, KARMEN and Bugey experiments, leaving only a little region
in parameter space.
The high $\delm$ region ($\delm > 4 eV^2$) can also be adressed by NOMAD. The
upgraded
ongoing KARMEN experiment as well as further data taking of LSND will clarify the
situation. Also the above mentioned reactor experiments may exclude some parts of
the allowed parameter region.
\section{Future accelerator experiments}
The next new results on \nosz from accelerators will come from CHORUS \cite{cn} and NOMAD
\cite{cn}
at CERN. Both are designed to improve the sensitivity for \nmu- \ntau - \oszs 
by at least one order of magnitude in the $\delm$-region for \hdm given by
the
quadratic see-saw-mechanism. Proposals for a next generation \nue detector at CERN exist,
which should be able to improve the sensitivity by another order of magnitude.
At Fermilab the proposed COSMOS-experiment
(P803) will reach roughly the same sensitivity.
The wish to go down to
smaller $\delm$ to check directly the region of
\atm requires either to go to larger
distances (long baseline experiments) and/or lower
energies.
The first experiment to happen will be an experiment 
shooting from KEK to Superkamiokande (KEK - E362). This experiment has a baseline
of 235 km
and is expected to start data taking beginning of 1999. An approved \lbl is the
MINOS-experiment using the Fermilab \nue beam to shoot to a detector in the
Soudan mine, about 730 km away. Roughly the same distance would apply for
proposed ideas for a CERN - Gran Sasso long baseline experiment. Several proposals
for detectors exist like ICARUS \cite{future}, NOE \cite{noe} and a 27 kt
Water-RICH \cite{tom}.
\section{Conclusions and outlook}
\begin{figure}[hhh]
\begin{center}
\epsfig{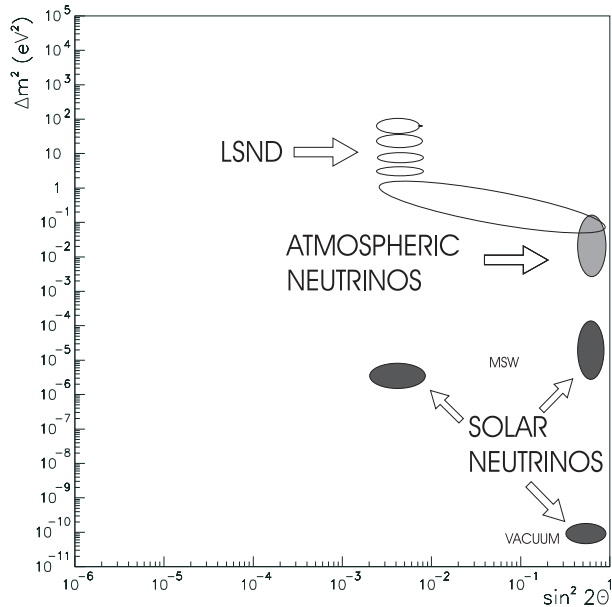}
\end{center}
\caption{\label{sum} \it Schematic presentation of prefered parameter regions to
explain the solar (dark), atmospheric (grey) and LSND (white) results with \noszp}
\end{figure}

Light neutrinos in the eV-region are still the best candidates for \hdmp While the
classical \ssm 
can be adjusted in a way that $\ntau$ is in the eV-range, the \adn scenarios allow
all of them as \dm candidates. Possible hints for \nue masses come out of solar
and
atmospheric neutrino data as well as the LSND experiment (fig.\ref{sum}).
\noindent
The main impact until the 
end of 1997 will come from Superkamiokande and further LSND measurements as well
as other experiments checking their result. Tab.2 
shows 
a comparison of different scenarios which can emerge out of these new
measurements. Depending on the special scenario different mass hierarchies and
models result and \nues of different flavour act as \dm candidate. 
Note that if both LSND and the atmospheric \nue problem are confirmed and
the solar neutrino data are included, it is not possible to explain all data with
the three standard model neutrinos. A possible solution could be additional
sterile neutrinos.
Only the scenario
with both atmospheric and LSND not confirmed, leaves room for a \ntau as \hdm
within the classical see-saw-mechanism. If LSND will be confirmed but not
atmospheric \nue problem, we even have an inverted mass hierarchy.
\medskip\\ 
\begin{center} 
{\large \begin{tabular}{|c|c|c|c|c|}
\hline solar & \multicolumn{4}{c|}{no dramatic new results} \\
\hline
atmos. & C & C & NC & NC \\
\hline
LSND & C & NC & C & NC \\
\hline
$\mid \Delta m^2_{12} \mid $ & $\approx 1$  & $\approx 10^{-5}$ & $\approx
1$&  
 $\approx 10^{-5}$ \\
\hline
$\mid \Delta m^2_{23} \mid $ & $\approx 10^{-2}$& $\approx 10^{-2}$  &  
 $\approx 1$& ? \\
\hline
$\mid \Delta m^2_{13} \mid $ & $\approx 1$& $\approx 10^{-2}$  & 
 $\approx 10^{-5}$ & ? \\
\hline
\hline
Short B.L. & $\nu_e - \nu_{\tau}$ & NO & YES  
& YES\\
\hline
Long B.L. & $\nu_\mu - \nu_{\tau}$ & YES & NO & 
 YES \\
\hline 
\hline 
HDM & $ \nu_{\mu,\tau}$ & $\nu_{e,\mu,\tau}$ & $ \nu_\mu $ &
$\nu_{\tau} (\nu_{e,\mu})$ \\
\hline
\end{tabular}}
\newline 
\end{center}
\medskip
{\it Tab. 2: Comparison of different scenarios depending on the
confirmation (C) or non-confirmation (NC) of the atmospheric neutrino data and
LSND results (partly compiled by L. diLella).
Also shown is the resulting observability in short and long-baseline experiments.
The \nues (eV-range) acting
as \hdm are shown in the last row. Only the last column allows the classical
see-saw mechanism.}
 
\end{document}